\documentclass[reprint,superscriptaddress,amsmath,amssymb,prl,showpacs,floatfix]{revtex4-1}
\usepackage{cases}
\usepackage{amsmath}
\usepackage{amssymb}
\usepackage{amsfonts}
\usepackage{amssymb}
\usepackage{dcolumn}
\usepackage{bm}
\usepackage{hyperref}
\usepackage{graphicx}
\usepackage{upgreek}
\usepackage{subfigure}
\usepackage{color}
\UseRawInputEncoding
\usepackage[columnwise]{lineno}
\begin{document}
	
	\title{Picotesla magnetometry of microwave fields with diamond sensors}
	
	\author{Zhecheng Wang}
	\affiliation{CAS Key Laboratory of Microscale Magnetic Resonance and School of Physical Sciences, University of Science and Technology of China, Hefei 230026, China}
	\affiliation{CAS Center for Excellence in Quantum Information and Quantum Physics, University of Science and Technology of China, Hefei 230026, China}
	
	\author{Fei Kong}
	\email{kongfei@ustc.edu.cn}
	\affiliation{CAS Key Laboratory of Microscale Magnetic Resonance and School of Physical Sciences, University of Science and Technology of China, Hefei 230026, China}
	\affiliation{CAS Center for Excellence in Quantum Information and Quantum Physics, University of Science and Technology of China, Hefei 230026, China}
	
	\author{Pengju Zhao}
	\affiliation{CAS Key Laboratory of Microscale Magnetic Resonance and School of Physical Sciences, University of Science and Technology of China, Hefei 230026, China}
	\affiliation{CAS Center for Excellence in Quantum Information and Quantum Physics, University of Science and Technology of China, Hefei 230026, China}
	
	\author{Zhehua Huang}
	\affiliation{CAS Key Laboratory of Microscale Magnetic Resonance and School of Physical Sciences, University of Science and Technology of China, Hefei 230026, China}
	\affiliation{CAS Center for Excellence in Quantum Information and Quantum Physics, University of Science and Technology of China, Hefei 230026, China}
	
	\author{Pei Yu}
	\affiliation{CAS Key Laboratory of Microscale Magnetic Resonance and School of Physical Sciences, University of Science and Technology of China, Hefei 230026, China}
	\affiliation{CAS Center for Excellence in Quantum Information and Quantum Physics, University of Science and Technology of China, Hefei 230026, China}
	
	\author{Ya Wang}
	\affiliation{CAS Key Laboratory of Microscale Magnetic Resonance and School of Physical Sciences, University of Science and Technology of China, Hefei 230026, China}
	\affiliation{CAS Center for Excellence in Quantum Information and Quantum Physics, University of Science and Technology of China, Hefei 230026, China}
	\affiliation{Hefei National Laboratory, Hefei 230088, China}
	
	\author{Fazhan Shi}
	\email{fzshi@ustc.edu.cn}
	\affiliation{CAS Key Laboratory of Microscale Magnetic Resonance and School of Physical Sciences, University of Science and Technology of China, Hefei 230026, China}
	\affiliation{CAS Center for Excellence in Quantum Information and Quantum Physics, University of Science and Technology of China, Hefei 230026, China}
	\affiliation{Hefei National Laboratory, Hefei 230088, China}
	\affiliation{School of Biomedical Engineering and Suzhou Institute for Advanced Research, University of Science and Technology of China, Suzhou 215123, China}
	
	\author{Jiangfeng Du}
	\email{djf@ustc.edu.cn}
	\affiliation{CAS Key Laboratory of Microscale Magnetic Resonance and School of Physical Sciences, University of Science and Technology of China, Hefei 230026, China}
	\affiliation{CAS Center for Excellence in Quantum Information and Quantum Physics, University of Science and Technology of China, Hefei 230026, China}
	\affiliation{Hefei National Laboratory, Hefei 230088, China}
	
	\begin{abstract}
		Developing robust microwave-field sensors is both fundamentally and practically important with a wide range of applications from astronomy to communication engineering. The Nitrogen-Vacancy (NV) center in diamond is an attractive candidate for such purpose because of its magnetometric sensitivity, stability and compatibility with ambient conditions. However, the existing NV center-based magnetometers have limited sensitivity in the microwave band. Here we present a continuous heterodyne detection method that can enhance the sensor's response to weak microwaves, even in the absence of spin controls. Experimentally, we achieve a sensitivity of 8.9 pT$\cdot$Hz$^{-1/2}$ for microwaves of 2.9 GHz by simultaneously using an ensemble of $n_{\text{NV}} \sim 2.8\times10^{13}$ NV centers within a sensor volume of $4\times10^{-2}$ mm$^3$. Besides, we also achieve $1/t$ scaling of frequency resolution up to measurement time $t$ of 10000 s. Our method removes the control pulses and thus will greatly benefit the practical application of diamond-based microwave sensors.
	\end{abstract}
	
	\maketitle
	
	
	Improving the sensitivity of microwave-field detection could directly advance many modern applications, such as wireless communication \cite{Holl2017}, electron paramagnetic resonance \cite{ESRbook}, high-field nuclear magnetic resonance \cite{Moser2017}, and even astronomical observations \cite{Pastor2021}. Instead of conventional inductive detection, various quantum sensors have been developed in the past decades with enhanced capabilities. For instance, Rydberg atoms \cite{Jing2020}, atomic magnetometers \cite{Gerginov2019}, superconducting quantum interference devices (SQUIDs) \cite{Couedo2019}, and nitrogen-vacancy (NV) centers in diamond \cite{Chipaux2015,Wang2015,Shao2016,Joas2017,Stark2017,Horsley2018,Meinel2021} are highly sensitive to either the electric or magnetic field of microwaves. Among them, NV centers are distinguished by their unique properties including solid-state aspect and room-temperature compatibility, which are essential for on-chip detection \cite{Kim2019}, but suffer from relatively low sensitivity.
	
	By using NV ensembles, the diamond magnetometry has demonstrated sub-pT$\cdot$Hz$^{-1/2}$ sensitivity for low-frequency ($< 1$ MHz) fields \cite{Wolf2015}. But it quickly degrades to sub-$\mu$T$\cdot$Hz$^{-1/2}$ level when sensing high-frequency (GHz) fields \cite{Horsley2018}. The reason is that the high-sensitivity protocols based on either pulsed or continuous dynamical decoupling are failed in the high-frequency range due to limited driving power \cite{Taylor2008,London2013}. Although this problem has recently been resolved by pulsed Mollow absorption \cite{Joas2017,Meinel2021} and concatenated continuous dynamical decoupling on single NV centers \cite{Stark2017}, generalizing those sophisticated control pulses to a larger area of NV ensembles remains challenging, due to the requirement of both strong and inhomogeneous control fields. To remove those barriers, one can directly observe the absorption of microwave by NV centers and the subsequent spin transitions via measuring either optically detected magnetic resonance (ODMR) spectra \cite{Chipaux2015,Shao2016} or Rabi oscillations  \cite{Wang2015,Horsley2018}. However, the absorption becomes inefficient and even loses the first-order response to the microwave field when the corresponding Rabi frequency is smaller than the inhomogeneous transition linewidth $\Delta \nu$.
	
	Here, we propose a continuous heterodyne detection scheme to enhance the absorption of weak microwave fields by introducing a moderate (still much weaker than $\Delta \nu$) and slightly detuned reference microwave. Under the illumination of a 532 nm laser, the continuously applied reference microwave interferes with the signal microwave, resulting in an oscillation of the NV photoluminescence. The oscillation directly gives the two important information of the signal microwave, field strength and frequency. We perform the demonstration on an ensemble of $n_{\text{NV}} \sim 2.8\times10^{13} $ NV centers within an effective sensor volume of $4\times10^{-2}$ mm$^3$. The diamond sensor maintains linear response to weak microwave fields down to subpicotesla. Within a total measurement time $t$ of 1000 s, a microwave field of 0.28 pT is detectable, corresponding to a sensitivity of 8.9 pT$\cdot$Hz$^{-1/2}$. The frequency resolution scales as $1/t$ down to 0.1 mHz for $t=10000$ s. We note that the strength of reference microwave required is just $\sim$ 200 nT. The removal of sophisticated control pulses makes our method applicable to larger diamond sensors with further improved sensitivity. It also greatly benefits practical applications.
	
	The NV electron spin has a triplet ground state consisting of a bright state $|m_S = 0\rangle$ and two degenerate dark states $|m_S = \pm 1\rangle$ (denoting as $|0\rangle$ and $|\pm1\rangle$ hereinafter) with a zero-field splitting of $D = 2.87$ GHz. The degeneracy can be lifted by an external magnetic field $\mathbf{B}$. Without loss of generality, we assume the microwave of the form $b_1 \cos \omega t$ is resonant with $|0\rangle \leftrightarrow |1\rangle$ transition (Fig.~1a), and then treat the NV center as a two-level system. If the microwave field $b_1$ is strong enough, one can observe a Rabi oscillation between $|0\rangle$ and $|1\rangle$ with frequency $g = \gamma_{\text{NV}}b_1/\sqrt{2}$, where $\gamma_{\text{NV}}$ is the gyromagnetic ratio of the NV electron spin. As shown in Fig.~1b, the oscillation slows down with reducing $b_1$, and finally degrades to an exponential decay. In addition to the intrinsic longitudinal relaxation $\Gamma_1 = 1/T_1$, the weak microwave opens an extra relaxation channel between $|0\rangle$ and $|1\rangle$ with rate of $\Gamma_g = g^2/\Gamma_2$ \cite{Hall2016}, where $\Gamma_2 = 1/T_2^{*}$ is the dephasing rate. In the presence of continuous laser, $| 1 \rangle$ will be polarized to $|0\rangle$ with a rate of $\Gamma_{\text{p}}$, which competes with the relaxation. A simple rate equation can describe the evolution of NV states (see Methods). In short, the competition leads to an equilibrium state, where the population of $|0\rangle$ is
	\begin{equation}
		P_0^{\infty} = \frac{1}{2}+\frac{\Gamma_{\text{p}}}{2(\Gamma_{\text{p}} + \Gamma_1 + \Gamma_g)}.
		\label{equilibrium}
	\end{equation}
	The extra relaxation $\Gamma_g$ will result in decreased fluorescence (Fig.~1c). In the weak-field limit, i.e., $\Gamma_g \ll \Gamma_1$, this decrement is $\propto \Gamma_g \propto b_1^2$, which means the NV center only preserves second-order response to the microwave.
	
	\begin{figure}
		\centering \includegraphics[width=1\columnwidth]{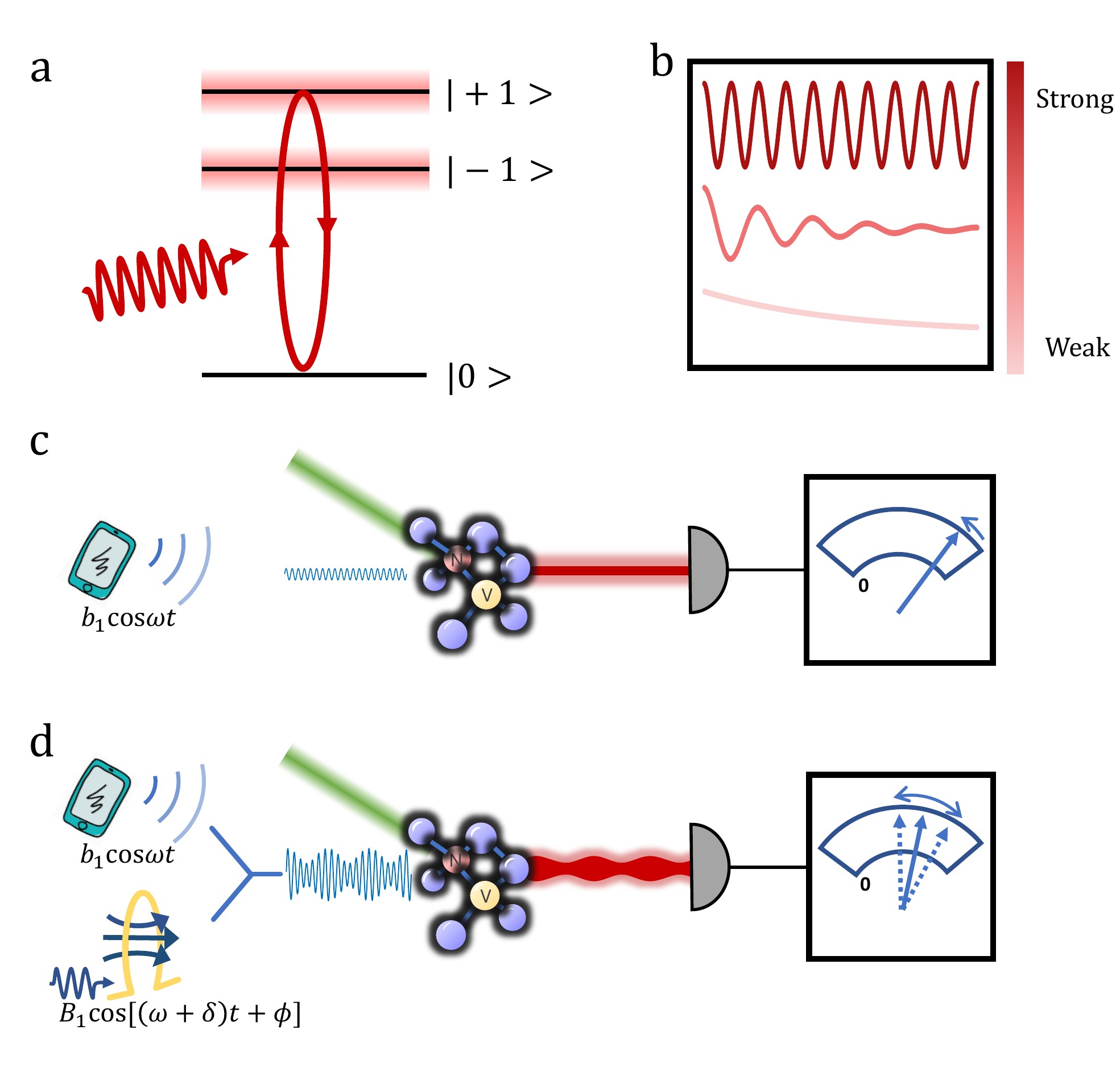} \protect\caption{
			\textbf{Basic principle of continuous heterodyne detection.}
			\textbf{a} Simplified energy levels of NV centers. The $|\pm 1\rangle$ states can be polarized to $|0\rangle$ state with a rate of $\Gamma_{\text{p}}$. A resonant microwave addresses the $| 0 \rangle \rightarrow | 1 \rangle$ spin transition. \textbf{b} Evolution of the NV center driven by microwaves of different magnitudes. For strong microwave, the spin state shows a Rabi oscillation between $|0\rangle$ and $|1\rangle$ with frequency $g$ proportional to the microwave magnitude. For weak microwave, the oscillation degrades to an exponential decay with rate proportional to the square of microwave magnitude. \textbf{c} Comparison of direct and heterodyne detection. The competition between laser-induced polarization and microwave-induced relaxation leads to a equilibrium spin state. For direct detection (upper), constant microwave magnitude results in DC fluorescence signal. For heterodyne detection (down), the microwave interference results in an time-varying magnitude, and thus an AC fluorescence signal.
		}
		\label{setup}
	\end{figure}
	
	If applying a reference microwave of the form $B_1 \cos [(\omega + \delta)t+\phi]$ simultaneously, a microwave interference will happen resulting in a modulation of the amplitude with beat frequency $\delta$, as shown in Fig.~1d. In the situation of $b_1 \ll B_1$, the amplitude can be simplified as $\sqrt{B_1^2 + 2 B_1 b_1 \cos(\delta t +\phi)}$. Here we assume both microwaves are resonant with the NV center, i.e. $\delta \ll \Gamma_2$, and thus the two microwaves can be treated as a single one leading to a time-varying relaxation rate of $\Gamma_G + 2\sqrt{\Gamma_G \Gamma_g} \cos(\delta t +\phi)$, where $\Gamma_G = G^2/\Gamma_2$ and $G = \gamma_{\text{NV}}B_1/\sqrt{2}$. The constant term $\Gamma_G$ induces a constant decrement of fluorescence, while the oscillating term induces an oscillation of fluorescence. The latter is predicted by the solution of rate equation (see Methods)
	\begin{equation}
		\Delta P_0^{\infty} (t) = \frac{\Gamma_{\text{p}} \sqrt{\Gamma_G \Gamma_g}\cos(\delta t +\varphi)}{(\Gamma_{\text{p}} + \Gamma_1 + \Gamma_G) \sqrt{(\Gamma_{\text{p}} + \Gamma_1 + \Gamma_G)^2 + \delta^2}}.
		\label{quasi_equilibrium} 
	\end{equation}
	The oscillation frequency is just the heterodyne frequency $\delta$, and the amplitude is $\propto \sqrt{\Gamma_g} \propto b_1$. Now the NV center has linear response to the microwave. We note the relaxation between $|0\rangle$ and $|-1\rangle$ will no longer negligible when $\Gamma_{\text{p}}$ is comparable or smaller than $\Gamma_1$. Nevertheless, the linear response does not change but with a slightly different coefficient (see SI).
	
	Benefit from the removal of complicated control pulses, we can perform the experiment on a simply-built setup (Fig.~2a). We use an optical compound parabolic concentrator (CPC) to enhance the fluorescence collection efficiency \cite{Wolf2015}. As a proof-of-principle demonstration, both the signal and reference microwaves are radiated from a 5-mm-diameter loop antenna. We apply an external magnetic field ($\sim$ 12.5 G) perpendicular to the diamond surface, so that all the NV centers have the same Zeeman splittings. A common ODMR spectrum (Fig.~2b) can be obtained by sweeping microwave frequency. The triplet feature arises from the hyperfine coupling between the NV electron spin and the $^{14}$N nuclear spin. The ODMR linewidth $\Delta \nu$ of 482 kHz is defined by the full width at half maximum (FWHM). Here both the laser and microwave are weak enough, so that $\Delta \nu \approx 2\Gamma_2$ \cite{Dreau2011}.
	
	\begin{figure*}
		\centering \includegraphics[width=1.8\columnwidth]{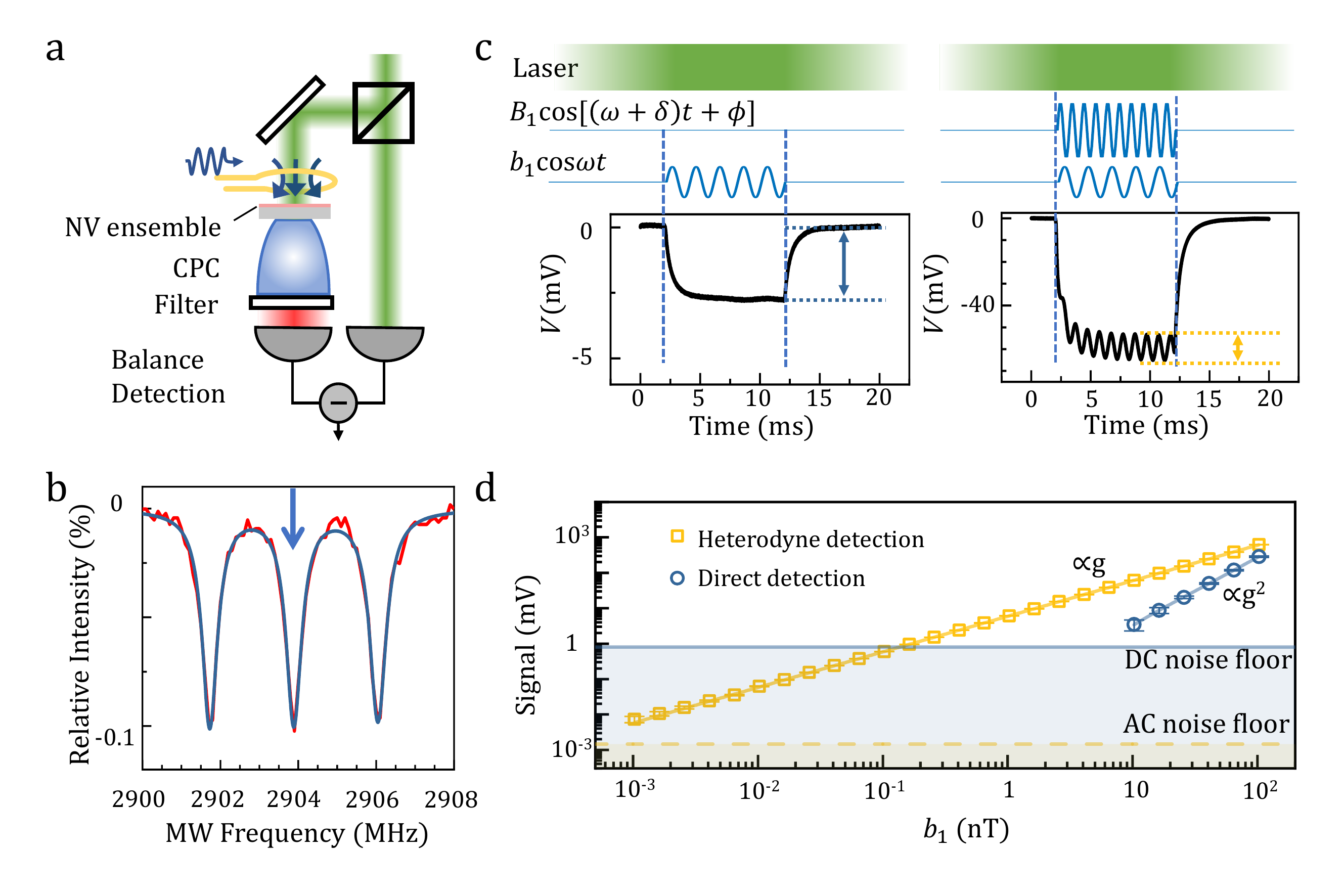} \protect\caption{
			\textbf{Proof-of-principle measurements on diamond sensors.}
			\textbf{a} Schematic of the setup. Both microwaves are radiated by a loop antenna with diameter of 5 mm, which is parallel with the diamond surface. \textbf{b} ODMR spectrum of the NV centers. The red line is the experimental result, while the blue line is three-peak Lorentz fit. The fitted FWHM linewidth is 482 kHz. The microwave field is 365 nT. The arrow marks the resonant frequency used in the following experiments. \textbf{c} Time trace of the photovoltage $V$ with continuous laser and pulsed microwaves. For single microwave (left), the photovoltage decreases to a constant value when the microwave is turned on, and revives after the microwave is turned off. The voltage difference marked by blue arrow is the signal strength of direct detection. For dual microwaves (right), the phenomenon is similar but shows an additional oscillation when the microwave is turned on. The oscillation amplitude marked by yellow arrow is the signal strength of heterodyne detection. Here $b_1 = 36.6$ nT, $B_1 = 201$ nT, and $\delta = 1$ kHz. \textbf{d} Dynamical range of the diamond sensor. Blue squares indicate the means of measured voltage differences, where error bars indicate SEM. Yellow circles are extracted from the Fourier transform spectra, where error bars indicate the RMS of base line around $\delta$ with a span of 0.1 Hz. The lines are linear and parabolic fits for direct and heterodyne detection, respectively. Blue and yellow areas indicate the DC and AC noise floor for a total measurement time of 1000 s.  
		}
		\label{setup}
	\end{figure*}
	
	We first apply a single-channel resonant microwave. As shown in Fig.~2c, the photovoltage begins to decrease from $V_0$ when the microwave is turned on, and approaches to a saturation value $V_{\infty}$. When the microwave is turned off, the photovoltage returns to $V_0$ with a slower revival rate of $\Gamma_1 + \Gamma_{\text{p}}$. Measurements of the dependence of the revival rate on the laser power $P_{\text{L}}$ (W) gives $\Gamma_1 = 102$ Hz and $\Gamma_{\text{p}} = 250 \times P_{\text{L}}$ Hz (see SI), where $P_{\text{L}} \leqslant 1.2$ W is much lower than the saturation power. As predicted by Eq.~\ref{equilibrium}, the voltage difference $\Delta V = V_0 - V_{\infty}$ shows squared dependence on the microwave field $b_1$ (Fig.~2d), and quickly lost in the noise. Here $b_1$ is calibrated according to the relation $g \propto \sqrt{P_{\text{MW}}}$, where $P_{\text{MW}}$ is the microwave power (see SI). Within a total measurement time of 1000 s, the minimum detectable $b_1$ is just $4.9$ nT. It means this direct measurement is insufficient to sensing weak fields. 
	
	We then apply a dual-channel resonant microwave. The time trace of photovoltage shows a similar behavior of decrease and revival with microwave turning on and off. Besides, it also shows an oscillation (Fig.~2c), where the frequency $f$ and amplitude $A$ can be directly extracted from the Fourier transform spectrum (not shown). $f$ is just the heterodyne frequency $\delta$ of the two microwaves, while $A$ is proportional to the signal field $b_1$. As shown in Fig.~2d, this linear response preserves over 5 orders in amplitude. Therefore, we have constructed a well field-to-voltage sensor with a dynamical range from 1 pT to 100 nT.
	
	To optimize the sensor's performance, we need improve the signal-to-noise ratio (SNR). According to Eq.~\ref{quasi_equilibrium}, the sensor's responsivity depends on multiple parameters $\Gamma_G$, $\Gamma_{\text{p}}$, $\delta$, and $\Gamma_1$, where the last is intrinsic and nonadjustable. We first focus on the reference microwave field $B_1$ (i.e. $\Gamma_G$), which is the key to enhance the response of NV centers to weak fields, serving as an amplifier. As shown in Fig.~3a, the signal first grows linearly with increasing $B_1$, then tends to saturate, and eventually goes down. Although this trend is consistent with the theoretical expectation, the specific experimental values still show obvious deviations from either Eq.~\ref{quasi_equilibrium} or the more accurate three-level model (SI). There are two reasons. First, the estimation of dephasing rate from the ODMR linewidth ($\Gamma_2 \approx \Delta \nu/2$) is inaccurate, where the static or quasi-static noise, such as local inhomogeneous strain fields or couplings to $^{13}$C nuclear spins, make considerable contribution. However, this kind of noise is not captured by the master equation (see SI). It will induce static or quasi-static shift of the resonant frequency, similar to the $^{14}$N hyperfine coupling, rather than speed up the transverse relaxation. Therefore, both the effective transverse relaxation rate $\Gamma_2$ and the effective contrast $C$ are overestimated. It can explain the optimal $B_1$ is smaller than the expectation. By loosing $\Gamma_2$ as a free parameter, the experimental results can be fitted by Eq.~\ref{quasi_equilibrium} better, where $\Gamma_{2,\text{fit}} = 152$ kHz. Second, $\Gamma_2$ is inhomogeneous for different NV centers, which can explain the experimental curve is more flat than the fit. More accurate models are unavailable without the prior knowledge of actual distribution of $\Gamma_2$. The optimal $B_1$ appears at 220 nT, corresponding to 4.35 kHz, well below the ODMR linewidth. We note the removal of strong control fields is the most important advantage of our method.
	
	\begin{figure}
		\centering \includegraphics[width=1\columnwidth]{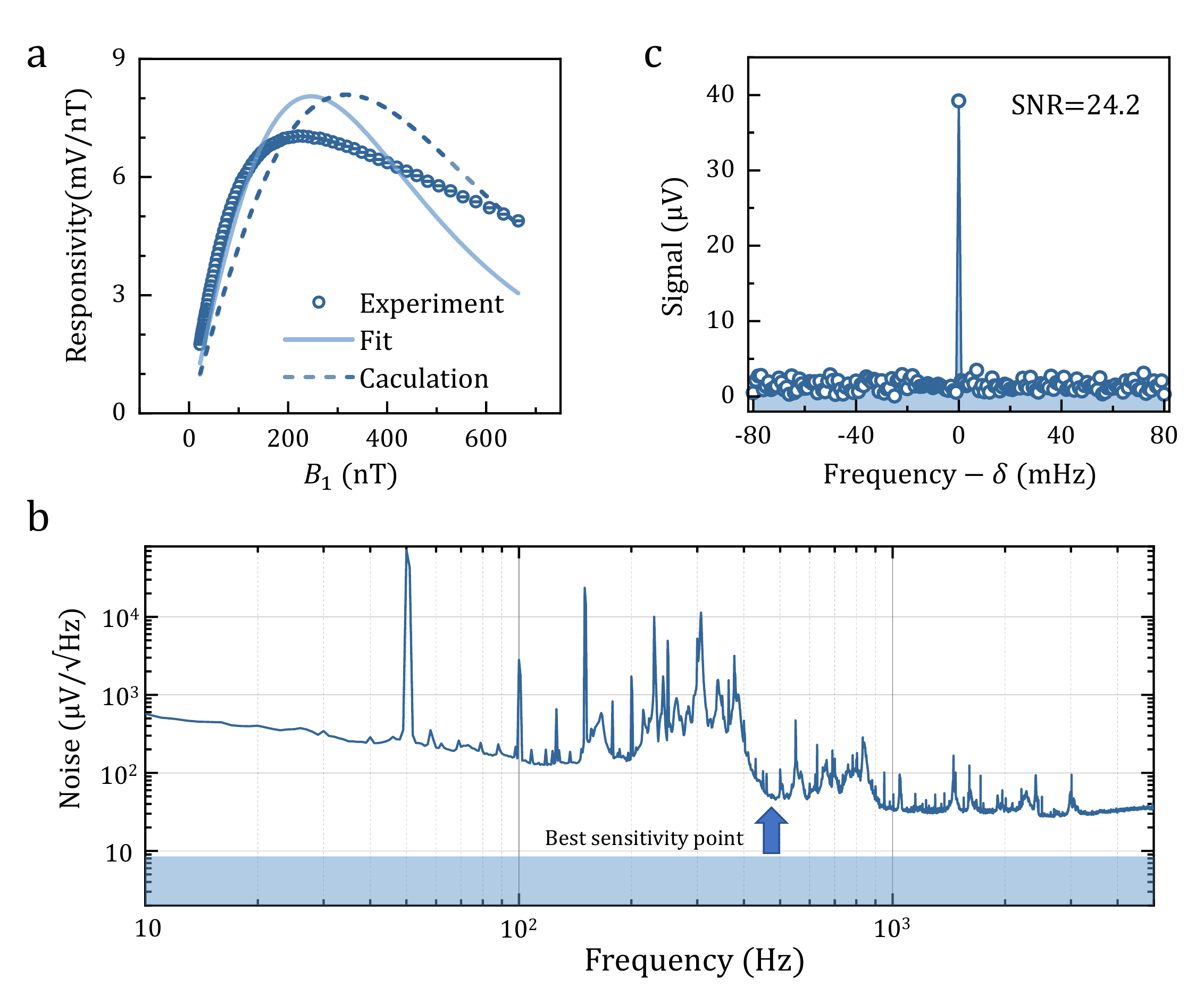} \protect\caption{
			\textbf{Optimal sensitivity.}
			\textbf{a} Dependence of responsivity on reference microwave field. Points are experimental results, where error bars indicate the RMS of base line in Fourier transform spectra around $\delta=480$ Hz with a span of 0.1 Hz. The dot line is theoretical calculations according to Eq.~\ref{quasi_equilibrium} with fixed $\Gamma_2 = 241$ kHz, while the solid line is fit with flexible $\Gamma_2$, and $\Gamma_{2,\text{fit}} = 144$ kHz. \textbf{b} Noise spectrum of the measurements. The noise trends to decrease with increasing $\delta$. The signal is also negatively correlated with $\delta$. The arrow marks the best SNR points at 480 Hz. The blue area indicate the estimated photon shot noise. \textbf{c} Benchmark of the sensitivity. The Fourier transform spectrum corresponds to a signal microwave field of 6.81 pT. The reference microwave field is 220 nT with $\delta = 480$ Hz. The total measurement time is 1000 s. The measured SNR of 24.2 corresponds to a sensitivity of 8.9 pT$\cdot$Hz$^{-1/2}$.  
		}
		\label{setup}
	\end{figure}
	
	Another adjustable parameter is the laser power $P_{\text{L}}$, which determines the spin polarization rate $\Gamma_{\text{p}} \propto P_{\text{L}}$. In principle, larger $P_{\text{L}}$ is preferred to maximal SNR if the noise is dominated by photon shot noise (see methods). However, the increase of $P_{\text{L}}$ will introduce more experimental imperfections, such as sample heating and laser-induced photon noise. Therefore, we choose a moderate laser power $P_{\text{L}} = 0.8$ W (see SI), corresponding to $\Gamma_{\text{p}} = 204$ Hz. At this point, the temperature of the diamond increases by 31.4 K. Figure~3b shows the noise spectral density of the fluorescence, which is contributed by the laser-induced noise, the electric noise of the photodetector, and photon shot noise. The first and last terms are proportional to the photon numbers $N$ and $\sqrt{N}$, respectively, while the second term is independent of $N$. Here the fluorescence is strong enough, so that only the laser-induced noise dominates. 
	
	For a constant-frequency signal microwave, we can adjust $\delta$ by tuning the frequency of the reference microwave. Due to the limited bandwidth, as we will discuss latter, increasing $\delta$ will reduce the signal, but it also brings lower noise (Fig.~3b). To maximal SNR, the optimal $\delta$ here is $\sim$ 480 Hz. Finally, we measure a microwave with frequency of 2903.9 MHz and field strength of 6.81 pT. Within a total measurement of 1000 s, the measured SNR is 24.2 (Fig.~3c), corresponding to a sensitivity of 8.9 pT$\cdot$Hz$^{-1/2}$.
	
	In addition to the high sensitivity, another highlight of our method is the unlimited frequency resolution, similar to previous heterodyne measurements \cite{Schmitt2017,Boss2017,Meinel2021}. As shown in Fig.~4a, the frequency resolution can be improve to 0.1 mHz by extending the total measurement time to 10000 s. Moreover, the frequency resolution do not show obvious deviations from the $1/t$ scaling up to 10000 s (inset of Fig.~4a). Another important figure-of-merit associated with frequency resolution is the bandwidth. According to Eq.~\ref{quasi_equilibrium}, the $-$3-dB bandwidth is $\sqrt{3}(\Gamma_{\text{p}} + \Gamma_1 + \Gamma_G)$, which is on the order of 100 Hz. A intuitive picture to understand this bandwidth is that the oscillation comes from the interference between the signal and reference microwave, but the fluorescence has limited-speed response to variations of the microwave field (Fig.~2c). The limited respond in time domain corresponds to the limited bandwidth in frequency domain. The diamond sensor thus serves like a narrow-band mixer. To extend bandwidth, we can cascade multiple `mixers', as shown in Fig.~4b. By simultaneously applying 240 channels of reference microwaves with frequency interval of 2 kHz, the `bandwidth' extends to 190 kHz, which is limited by the ODMR linewidth. Although the sensor has response to all microwaves within the bandwidth, the measured frequency is the frequency difference between the signal microwave and the nearest reference microwave. It means frequency aliasing will happen, and thus we need repeat the measurement with different frequency interval of the reference microwaves to extract the actual frequency. We note the extension of bandwidth comes at the expense of sensitivity. When the signal microwave is interfering with one of the reference microwaves, the others serves like noise sources inducing additional longitudinal relaxation, corresponding to larger $\Gamma_1$, and thus poorer sensitivity. Specifically, for $m$ channels of reference microwaves, the bandwidth scales as $m$, while the sensitivity scales as $\sqrt{m}$ (see Methods).
	
	\begin{figure}
		\centering \includegraphics[width=1\columnwidth]{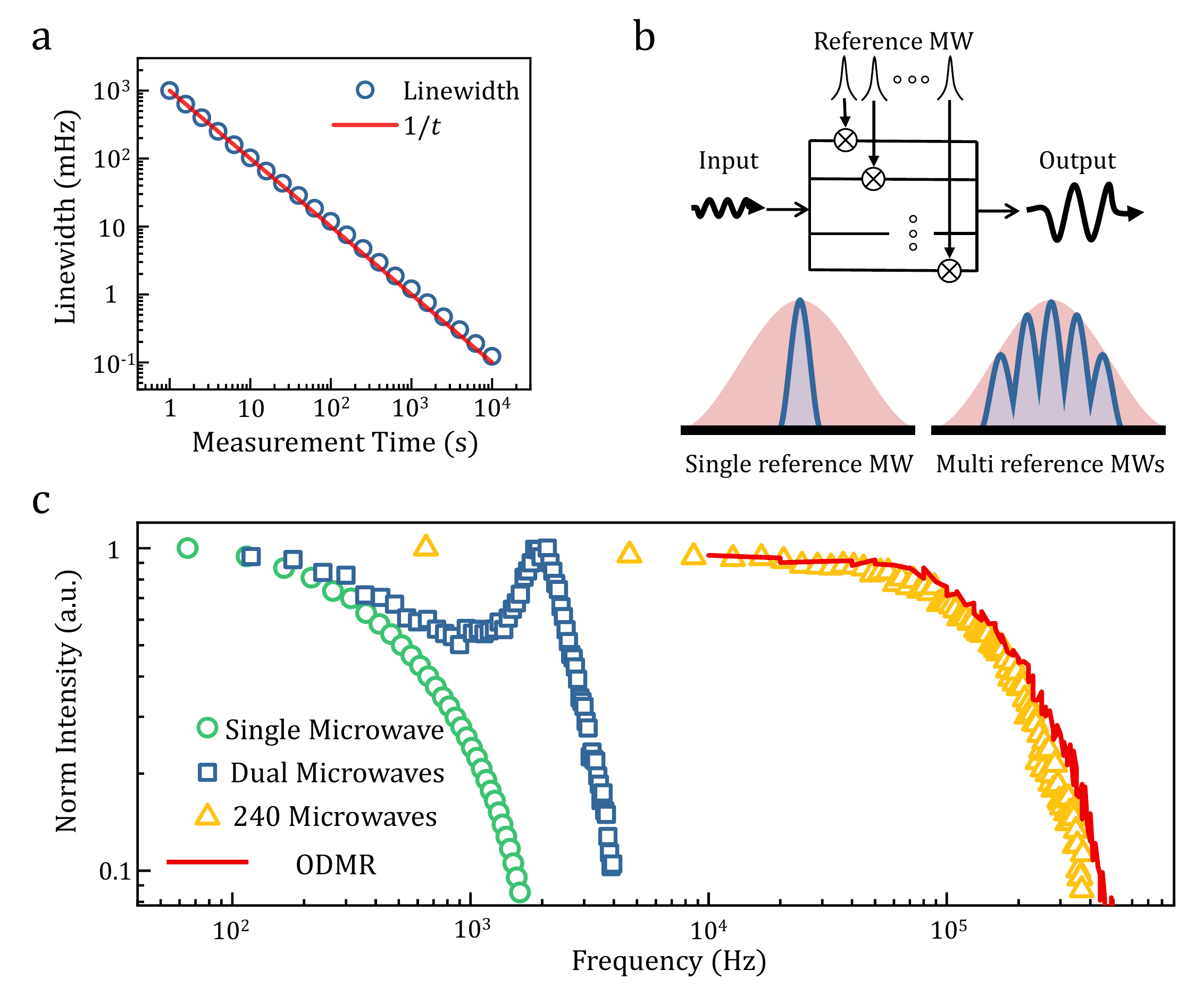} \protect\caption{
			\textbf{Linewidth and Bandwidth.}
			\textbf{a} Fourier transform spectrum of a signal measured for 10000 s. The FWHM linewidth extracted from Lorentz fit is 0.1 mHz. Inset gives the dependence of linewidth on the total measurement time $t$, where the line indicates $\propto 1/t$ scaling. \textbf{b} Intuitive concept of bandwidth extension. The diamond `mixer' has a narrow-band response to the input microwave, where the band is centered at the frequency of reference microwave. If cascading multiple mixers with different reference microwaves, the band will be the extended accordingly. \textbf{c} Measurements of bandwidth. All groups of measurement are normalized for better comparison of the bandwidth. The extended bandwidth consists with the ODMR linewidth. 
		}
		\label{setup}
	\end{figure}
	
	\textbf{Conclusions} We have shown that NV centers can be used as highly sensitive sensors for microwave magnetometry even in the absence of spin controls. Our method is based on the resonant absorption of microwaves by NV centers. We significantly improve its response to weak fields by introducing a moderate reference microwave. Furthermore, we applied the method to a diamond containing ensembles of NV centers, achieving minimum detectable microwave field of 0.28 pT and frequency resolution of 0.1 mHz. 
	
	Benefit from the simplicity of our method, the measurements can be directly reproduced on larger sensors, resulting further improved sensitivity. For example, if the diamond has a similar size as the photodiode ($\sim 10\times 10 \times 1$ mm$^3$), the sensitivity can be directly promoted to fT level. Even so, the sensor is still far smaller than microwave wavelengths. An increase in NV density will also leads to improved sensitivity, but needs to balance the increase of relaxation rate and the accompanying laser heating problem. The current submegahertz detection bandwidth is limited by the ODMR linewidth. Improvement to gigahertz is possible by introducing a magnetic field gradient \cite{Chipaux2015} and more reference microwaves. On the other hand, the sensitive frequency band is determined by the spin transition, which is magnetically adjustable up to hundreds of GHz \cite{Fortman2021} and even to the THz range in the future. Another small-range tuning method is using the Floquet dressed states that are independent of transition frequency \cite{Meinel2021}. 
	
	Our work paves the way for practical applications of diamond sensors, for example, microwave receivers in radars \cite{Lloyd2008}, wireless communications \cite{Koenig2013}, and even radio telescopes \cite{Pastor2021}. This diamond devices can work in extreme conditions, such as high temperature \cite{Liu2019}, high pressure \cite{Hsieh2019,Yip2019,Lesik2019}. The removal of spin controls will also efficiently reduce the complexity of contracting an on-chip diamond magnetometer \cite{Kim2019}.
	
	\section{MATERIALS AND METHODS}
	\subsection{Experimental setup and diamond samples}
	
	The core of our experimental setup is a sandwich of diamond, CPC (Edmund 45DEG2.5MM) and optical filter (Semrock LP02-638RU-25), which is mounted on a photodiode (Thorlabs PDAPC2) for fluorescence detection. A high-power laser (Lighthouse Sprout-D-5W) is used for illumination. The microwaves are generated from two RF signal generators (Stanford SG386) and/or an arbitrary waveform generator(Keysight M8190a), amplified by a microwave amplifier (Mini-circuits ZHL-25W-63+) or attenuated by several attenuators, and radiated by a home-build loop antenna with a diameter of 5 mm. Another photodiode (Thorlabs PDA36A2) is used for laser detection. The photovoltage is detected by an oscilloscope (NI PXIe-5122) after a differential amplifier (Stanford SR560). For the 10000 s measurement (Fig.~4a), another oscilloscope (Keysight MSOS254A) is used. All clocks are synchronized by one of the RF signal generators. The diamond is 100-oriented with an extra high-doping layer $\sim$ 10 $\mu$m thick growth with 99.99\% $^{12}$C isotopic purity. The effective sensor volume is $\sim4\times10^{-2}$ mm$^{3}$ with estimated NV density of $\sim$ 4 ppm. So the total number of NV centers is $n_{\text{NV}} \sim 2.8\times10^{13}$. 
	
	\subsection{Rate equations}
	Considering a two-level system consisting of $|0\rangle$ and $|1\rangle$, there exists a relaxation between them of the rate $\Gamma_1 + \Gamma_{g}$ and a polarization from $|1\rangle$ to $|0\rangle$ of the rate $\Gamma_{\text{p}}$. So the rate equations are
	\begin{equation}
		\begin{split}
			P_0^{'} & = -\frac{\Gamma_1 + \Gamma_g}{2}(P_0 - P_1) + \Gamma_{\text{p}}P_{1} \\	
			P_1^{'} & = -\frac{\Gamma_1 + \Gamma_g}{2}(P_1 - P_0) - \Gamma_{\text{p}}P_{1},
		\end{split}
		\label{rate_equation}
	\end{equation}
	where $P_j$ is the population of $|j\rangle$, and satisfies $\sum P_j \equiv 1$. The solution is
	\begin{equation}
		P_0(t) = P_0^{\infty} + [P_0(0) - P_0^{\infty}]e^{-(\Gamma_1 + \Gamma_g + \Gamma_{\text{p}})t},
		\label{solution_rate_equation}
	\end{equation}
	where $P_0^{\infty}$ is the equilibrium solution ($P_j^{'}=0$) given by Eq.~\ref{equilibrium}. If applying two microwaves, the rate equations becomes
	\begin{equation}
		\begin{split}
			P_0^{'} & = -\frac{\Gamma_1 + \Gamma_G + 2\sqrt{\Gamma_G \Gamma_g} \cos(\delta t +\phi)}{2}(P_0 - P_1) + \Gamma_{\text{p}}P_{1} \\	
			P_1^{'} & = -\frac{\Gamma_1 + \Gamma_G + 2\sqrt{\Gamma_G \Gamma_g} \cos(\delta t +\phi)}{2}(P_1 - P_0) - \Gamma_{\text{p}}P_{1}.
		\end{split}
		\label{rate_equation2}
	\end{equation}
	Considering that $\Gamma_g \ll \Gamma_G$, the time-varying term is only a perturbation of the relaxation rate, so the solution $P_0^{\infty}(t)$ should also be a perturbation $\Delta P_0^{\infty}(t)$ of the equilibrium solution $P_0^{\infty}$. Substituting a trail solution $P_0^{\infty}(t) = P_0^{\infty} + A \cos(ft+ \varphi)$ into Eq.~\ref{rate_equation2} yields $f = \delta$, $\varphi = \phi + \arctan[\delta/(\Gamma_{\text{p}} + \Gamma_1 + \Gamma_G)]+\pi$, and
	\begin{equation}
		A = \frac{\Gamma_{\text{p}} \sqrt{\Gamma_G \Gamma_g}}{(\Gamma_{\text{p}} + \Gamma_1 + \Gamma_G) \sqrt{(\Gamma_{\text{p}} + \Gamma_1 + \Gamma_G)^2 + \delta^2}}.
	\end{equation}
	
	\subsection{Shot-noise-limited sensitivity}
	If the laser power is much lower than the saturation power, the photon emission rate of NV centers is proportional to the polarization rate $\Gamma_{\text{p}}$. In a sample period $T_{\text{s}}$, the detected photons are $n_{\text{NV}}K\Gamma_{\text{p}}T_{\text{s}}$, where $K$ is constant depending on the collection efficiency. Therefore, within a total measurement time of $t$, the SNR of Fourier transform spectra is
	\begin{equation}
		\begin{split}
			\text{SNR} & = \frac{n_{\text{NV}}K\Gamma_{\text{p}}T_{\text{s}}\cdot C \cdot A}{2\sqrt{n_{\text{NV}}K\Gamma_{\text{p}}T_{\text{s}}}}\cdot \sqrt{\frac{t}{T_{\text{s}}}} \\
			& = \frac{\Gamma_{\text{p}}^{\frac{3}{2}} \Gamma_G^{\frac{1}{2}} \cdot C\sqrt{n_{\text{NV}}K\Gamma_g t}}{2(\Gamma_{\text{p}} + \Gamma_1 + \Gamma_G) \sqrt{(\Gamma_{\text{p}} + \Gamma_1 + \Gamma_G)^2 + \delta^2}} \\
			& < \frac{\Gamma_{\text{p}}^{\frac{3}{2}} \Gamma_G^{\frac{1}{2}}}{2(\Gamma_{\text{p}} + \Gamma_1 + \Gamma_G)^2} \cdot C\sqrt{n_{\text{NV}}K\Gamma_g t},
		\end{split}
		\label{SNR}
	\end{equation} 
	where $C$ is the fluorescence contrast of NV centers. The noise is defined by the RMS of base line in Fourier transform spectra, so there exists a coefficient 2 in the denominator. Here one can see that larger $\Gamma_{\text{p}}$ and $\Gamma_G $ are preferred, and the SNR quickly saturates to $\frac{3\sqrt{3}}{32}C\sqrt{n_{\text{NV}}K\Gamma_g t}$ when $\Gamma_{\text{p}}=3\Gamma_G > \Gamma_1$. Therefore, the shot-noise-limited sensitivity is
	\begin{equation}
		b_{\text{min}} > \frac{32\sqrt{2}}{3\sqrt{3}C} \sqrt{\frac{\Gamma_2}{n_{\text{NV}}Kt}}.
	\end{equation}
	For $m$ channels of reference microwaves, only one of them interference with a specific single microwave and others induced additional longitudinal relaxation. So Eq.~\ref{SNR} becomes
	\begin{equation}
		\text{SNR}  < \frac{\Gamma_{\text{p}}^{\frac{3}{2}} \Gamma_G^{\frac{1}{2}}}{2(\Gamma_{\text{p}} + \Gamma_1 + m\Gamma_G)^2} \cdot C\sqrt{n_{\text{NV}}K\Gamma_g t}.
	\end{equation} 
	It saturates to $\frac{3\sqrt{3}}{32\sqrt{m}}C\sqrt{n_{\text{NV}}K\Gamma_g t}$ when $\Gamma_{\text{p}}=3m\Gamma_G > \Gamma_1$.
	
	\section*{Acknowledgments}
	We thank Liang Zhang for helpful discussions. We thank element six for providing the diamond. \textbf{Funding:} This work was supported by the National Natural Science Foundation of China (Grants Nos. 81788101, 31971156, T2125011), the National Key R\&D Program of China (Grants Nos. 2018YFA0306600), the CAS (Grants Nos. XDC07000000, GJJSTD20200001, and YIPA2015370), and the Anhui Initiative in Quantum Information Technologies (Grant No. AHY050000). \textbf{Author contributions:} J.D. and F.S. supervised the entire project. F.K. and F.S. designed the experiments. Z.W., F.K., and P.Z. prepared the setup and performed the experiments. F.K. and Z.H. carried out the calculations. P.Y. and Y.W. prepared the test diamonds. F.K., Z.W., and F.S. wrote the manuscript. All authors discussed the results and commented on the manuscript. \textbf{Competing interests:} The authors declare no competing interests. \textbf{Data and materials availability:} All data needed to evaluate the conclusions in the paper are present in the paper and/or the Supplementary Materials. Additional data related to this paper may be requested from the authors.
	
	\renewcommand\refname{Reference}

	
\end{document}